\documentclass[preprint,12pt,english]{elsarticle}
\usepackage[english]{babel}
\usepackage{amsmath,amsthm,amsfonts,amssymb}
\usepackage[mathcal]{eucal}
\usepackage{mathrsfs}
\usepackage{natbib}
\biboptions{square,numbers,sort&compress}
\usepackage{lmodern}
\usepackage{graphicx}
\usepackage{dcolumn}
\usepackage{latexsym}
\usepackage{epsfig}
\usepackage{bm}
\usepackage[all]{xy}
\usepackage[utf8]{inputenc}
\usepackage{scalefnt}
\usepackage{dsfont}
\usepackage{hyperref}
\usepackage{braket}
\DeclareMathOperator{\e}{e}

\newcommand{\bb}{\text{b}}
\usepackage{adjustbox}
\usepackage{subfigure}
\usepackage{color}

\usepackage{resizegather}
\newcommand{\affiliation}{\address}

\begin{document}

%\begin{frontmatter}

\title{Quasi-normal modes of bumblebee wormhole}
\author{R. Oliveira$^{a}$}
\ead{rondinelly@fisica.ufc.br}

\author{D. M. Dantas$^{a}$}
\ead{davi@fisica.ufc.br}

\author{Victor Santos$^{a}$}
\ead{victor\_santos@fisica.ufc.br}

\author{C. A. S. Almeida$^{a}$}
\ead{carlos@fisica.ufc.br}

\affiliation{$^{a}$Universidade Federal do Cear\'a (UFC), Departamento de F\'{\i}sica, Campus do Pici, Caixa Postal 6030, 60455-760, Fortaleza, Cear\'{a}, Brazil}

%\keywords{Gravitational Waves, Reege-Wheeler's Potential,  Quasi-normal Modes}

\begin{abstract}

In this work, we calculate the quasi-normal frequencies from a bumblebee traversable wormhole. The bumblebee wormhole model is based on the bumblebee gravity, which exhibits a spontaneous Lorentz symmetry breaking. Supporting by the Lorentz violation parameter $\lambda$, this model allows the fulfillment of the flare-out and energy conditions, granted non-exotic matter to the wormhole. We analyze the parameters of bumblebee wormhole in order to obtain a Reege-Wheeler's equation with a bell-shaped potential. We obtain the quasi-normal modes (QNMs) via the WKB approximation method for both scalar and gravitational perturbations. All frequencies obtained are stable and the time domain profiles have decreasing oscillation (damping) profiles for the bumblebee wormhole. 
\end{abstract}
\maketitle

%\end{frontmatter}

\section{Introduction}

Wormholes are solutions of Einstein equation denoted by tunnels that connect two different regions of spacetime \cite{einstein-rosen, misner}.  The first conception of such structures comes from 1935 by the Einstein-Rosen bridge \cite{einstein-rosen}, being the term wormhole later adopted by Misner and Wheeler in 1957 \cite{misner}. Unlike black holes, wormholes have no horizon, and it can be traversable depending on some conditions at the wormhole throat \cite{flare, morris, morris2, small}. 

In a very recently work, Bueno \textit{et al.} \cite{echoes} present wormholes as a model for a class of exotic compact objects (ECOs). As a matter of fact, in the context of gravitational waves and its detections \cite{abb}, wormholes would be distinguished from black holes due to its features of ECOs, which can exhibit two bumps, instead of the usual single bump characteristic of black holes \cite{echoes}. On the other hand,  traversable wormholes violate the null convergence condition (NCC) close at the throat, which leads to a violation of the null energy condition (NEC) \cite{Energy1, non-exotic}. Hence, in the context of the usual formulation of Einstein-Hilbert General Relativity, traversable wormholes need an exotic matter source  \cite{non-exotic, fr-wormhole, qnm-wormhole, gb-wormhole}. However, recently some approaches were proposed in order to avoid such exotic matter by the modification of the gravity, namely, with the wormhole in a Born-Infeld gravity \cite{non-exotic}, in the $f(R)$ theories \cite{fr-wormhole}, in the framework of Gauss-Bonnet \cite{gb-wormhole} and considering the wormhole in the bumblebee gravity scenario \cite{Ovgun}.
 
The so-called bumblebee model was developed from the string theory, which a spontaneous Lorentz symmetry breaking (LSB) was verified \cite{kosu}. Such violations imply in the conception of the so-called Standard Model Extension (SME) \cite{coll, kost}. Recently, a Schwarzschild-like solution on a bumblebee gravity was proposed in Ref. \cite{casana}, where the Lorentz violation parameter is upper-bounded by the Shapiro's time delay of light. Moreover, a traversable wormhole solution in the framework of the bumblebee gravity was proposed by \"{O}vg\"{u}n  \textit{et al.} \cite{Ovgun}. In this bumblebee wormhole, the Lorentz violation parameter can support a normal matter source. At the limit of vanishing LSB, the deflection of light in this model becomes the same of the Ellis wormhole \cite{ellis}.

In this letter, we calculate the quasi-normal modes (QNMs) and time domain profiles for scalar and gravitational perturbations in the \"{O}vg\"{u}n bumblebee wormhole \cite{Ovgun}. We obtain a range over the Lorentz violation parameter $\lambda$ that satisfies to all flare-out and energy conditions. Furthermore, we also choose suitable LSB parameter that performs stable quasi-normal modes with a decreasing time-domain profile. 

The paper is organized as follows. In Sec \ref{sec-ovgun}, the \"{O}vg\"{u}n bumblebee wormhole is reviewed. In Sec \ref{sec-conditions}, the energy conditions and the flare-out conditions are analyzed by the LSB parameter. As a matter of fact, here we obtain a tranversable non-exotic wormhole. In subsection \ref{sec-tort}, we introduce the change of variable to obtain the Regge-Wheeler equation \cite{regge}. In Sec \ref{sec-rw}, we describe the Regge-Wheeler equation for the scalar and tensorial perturbations in the bumblebee wormhole, where we made a suitable choice of parameters to obtain an exact bell-shaped Regge-Wheeler potential. In subsection \ref{sec-qnm}, the quasi-normal frequencies were computed by the third-order  WKB method \cite{iyer} and the damping time-domain profile was evaluated for both scalar and tensorial perturbations.  In Sec \ref{sec-conclu}, we present our last discussions and summarize our results.

%=============================================================================================================================
\section{Bumblebee wormhole}\label{sec-ovgun}
%=============================================================================================================================

In this section, we review the exact solution of bumblebee wormhole presented in Ref. \cite{Ovgun}. Let us start with the  following bumblebee action
\begin{equation}
S_{B} = \int dx^{4} \sqrt{-g} \left[\frac{R}{2\kappa} + \frac{1}{2\kappa}\xi B^{\mu}B^{\nu} R_{\mu\nu} - \frac{1}{4} B_{\mu\nu}B^{\mu\nu} - V(B_{\mu}B^{\mu} \pm b^{2})\right] + \int dx^{4} \mathcal{L}_{m},
\end{equation}
where $B_{\mu}$ represents the bumblebee vector field, the $B_{\mu \nu}=\partial_{\mu}B_{\nu}-\partial_{\nu}B_{\mu}$ is the bumblebee field strength, and the $\xi$ is the non-minimal curvature coupling constant. For the vacuum solutions $V(B_{\mu}B^{\mu} \mp \bb^{2})=0$, we have that $\bb^2=\pm B^{\mu}B_{\mu}=\pm \bb^{\mu}\bb_{\mu}$ is the non-null vector norm associated to the vacuum expectation value $\braket{B^{\mu}}= \bb^{\mu}$ \cite{Ovgun,kost}. The scalar curvature is denoted by $R$, $g$ is the metric determinant and $\kappa$ the gravitational constant.

The energy-momentum tensor is modified by the bumblebee field in the following form \cite{casana, Ovgun}:
\begin{equation}
R_{\mu \nu }-\kappa G\left[ T_{\mu \nu }^{M}+T_{\mu \nu }^{B}-\frac{1}{2}%
g_{\mu \nu }\left( T^{M}+T^{B}\right) \right] =0, \label{einstein-1}
\end{equation}%
where $T^{M}=g^{\mu \nu }T_{\mu \nu }^{M}$ and the bumblebee energy-momentum tensor $T_{\mu\nu}^{B}$ reads
\begin{gather}
T_{\mu \nu }^{B}=-B_{\mu \alpha }B_{\nu }^{\alpha }-\frac{1}{4}B_{\alpha
\beta }B^{\alpha \beta }g_{\mu \nu }-Vg_{\mu \nu }+2V^{\prime }B_{\mu
}B_{\nu }+  \nonumber \\
+\frac{\xi }{\kappa }\Big[\frac{1}{2}B^{\alpha }B^{\beta }R_{\alpha \beta
}g_{\mu \nu }-B_{\mu }B^{\alpha }R_{\alpha \nu }-B_{\nu }B^{\alpha
}R_{\alpha \mu }+  \nonumber \\
+\frac{1}{2}\nabla _{\alpha }\nabla _{\mu }(B^{\alpha }B_{\nu })-\frac{1}{2%
}\nabla ^{2}(B_{\mu }B_{\nu })+  \nonumber \\
-\frac{1}{2}g_{\mu \nu }\nabla _{\alpha }\nabla _{\beta }(B^{\alpha
}B^{\beta })\Big]. \label{em-tensor}
\end{gather}%

The modified Einstein equation in Eq. \eqref{einstein-1} with the energy-momentum tensor in Eq. \eqref{em-tensor} can be explicited as
\begin{gather}
E_{\mu \nu }^{instein} = R_{\mu \nu }-\kappa \left( T_{\mu \nu }^{M}-\frac{1%
}{2}g_{\mu \nu }T^{M}\right) -\kappa T_{\mu \nu }^{B}-2\kappa g_{\mu \nu }V+ 
\nonumber \\
+\kappa B_{\alpha }B^{\alpha }g_{\mu \nu }V^{\prime }-\frac{\xi }{4}g_{\mu
\nu }\nabla ^{2}(B_{\alpha }B^{\alpha }) + \nonumber \\
-\frac{\xi }{2}g_{\mu \nu }\nabla _{\alpha }\nabla _{\beta }(B^{\alpha
}B^{\beta })=0.  \label{einstein-2}
\end{gather}%

At this point, the authors of Ref. \cite{Ovgun} choose a static and  spherically
symmetric traversable wormhole solution in the following form \cite{Ovgun, morris2}
\begin{equation}
ds^2 = e^{2\Lambda} dt^{2} - \frac{dr^{2}}{1 - \frac{b(r)}{r}} - r ^2 d\theta^{2} - r^2 \sin^{2}\theta d\phi^{2}, \label{metric}
\end{equation}
where the red-shift function is made null $(\Lambda=0)$ and the  bumblebee vector $\bb_{\mu}$ is set to be correlated to the  wormhole shape function $b(r)$ as following \cite{Ovgun}
\begin{equation}
\bb_{\mu} = \left(0, \sqrt{\frac{a}{1 - \frac{b(r)}{r}}}, 0,0 \right),
\end{equation}
where $a$ is a positive constant associated with the Lorentz violation term.

Besides, following the reference \cite{Ovgun}, the isotropic energy–momentum tensor can be decomposed as a perfect fluid  $ (T^{\mu}_{\nu})^{M} = (\rho, -P,-P,-P)$ where 
\begin{eqnarray}
P=w \rho, 
\end{eqnarray}
assuming $\rho \geq 0$. The $ -\frac{1}{3} < w \leq 1$ is a dimensionless constant responsible to hold the energy conditions.

Substituting the wormhole metric Eq. \eqref{metric} into Einstein equation \eqref{einstein-2}, we have the following new Einstein equations with the $b(r)$ and $w$ dependence
\begin{gather}
\label{einstein-a}
E_{t t }^{instein}= -\kappa \rho {r}^{3}\left(1+3w\right)+\lambda
r\dot{b}(r) -\lambda b(r)=0
,\\ \label{einstein-b}
E_{r r }^{instein}=  \kappa \rho {r}^{3}\left(w-1\right)+\left(2+3\lambda\right)r\dot{b}-\left(2+3\lambda\right)b(r)=0
,\\
E_{\theta \theta }^{instein}= \kappa \rho {r}^{3}\left(w-1\right)+r\dot{b}(r)+\left(2\lambda+1\right)b(r)-2\lambda r =0
,\label{einstein-c}
\end{gather}
where  $\lambda = a \xi$ is defined as the Lorentz symmetry breaking (LSB) parameter and the dot denotes derivative with respect to the coordinate $r$.

The energy density $\rho$ can be obtained from the system of equations $\left(\ref{einstein-a}-\ref{einstein-c}\right)$. By solving directly  $\rho$ from Eq. \eqref{einstein-a} we obtain:
\begin{equation}
\label{rho-a}
\rho = \frac{\lambda\left( r \dot{b}(r)  - b(r)\right)  }{\kappa r^{3} (1 + 3w)}.
\end{equation}

Moreover, $b(r)$ can be found by multiplying  $(w-1)$ by Eq. \eqref{einstein-a} and summing with  $(1+3w)$ multiplied by Eq. \eqref{einstein-c}. Additionally, the condition at wormhole throat $b(r_0)=r_0$ leads to the solutions
\begin{equation}
b(r) = \frac{\lambda r}{\lambda + 1} + \frac{r_{0}}{\lambda + 1} \left(\frac{r_{0}}{r} \right)^{\gamma}, \label{br}
\end{equation}
where $\gamma(w,\lambda)=\frac{\lambda (5 w+3)+3 w+1}{\lambda (w-1)+3 w+1}$.

The derivative $\dot{b}(r)$ can explicitly be obtained as
\begin{equation}
\dot{b}(r) = \frac{\lambda}{\lambda + 1}\left[1-\gamma\left(\frac{r_{0}}{r} \right)^{\gamma+1}\right]. \label{dbr}
\end{equation}

Finally, from Eq. \eqref{br} and Eq. \eqref{dbr}, the energy-density of Eq. \eqref{rho-a} can be found in the form
\begin{equation}
\label{rho-b}
\rho(r) =  - \frac{2 \lambda r_0^{\gamma+1}r^{-\left(\gamma+3\right)}}{\kappa \left(\lambda (w-1)+3 w+1\right)}
%\rho(r) =  - \frac{2 \lambda r_0^{\frac{2 (\lambda+1) (3 w+1)}{\lambda (w-1)+3 w+1}}r^{-\frac{4 ((2 \lambda+3) w+1)}{\lambda (w-1)+3 w+1}}}{\kappa \left(\lambda (w-1)+3 w+1\right)}\, .
\end{equation}

In next section, we analyze the parameters $\lambda$ and $w$ in order to obtain a special case of the \"{O}vg\"{u}n  wormhole, where all energy conditions and flare-out holds. Moreover, we will also set these parameters in order to obtain a Regge-Wheeler potential with a bell-shaped curve.

\section{Energy conditions and flare-out}\label{sec-conditions}

In this section we analyze the conditions imposed in the wormhole. We summarize the conditions in Table \ref{tab-conditions}. The NEC is the Null energy condition, WEC is the Weak energy conditions, SEC is the Strong energy condition, DEC is the Dominant energy conditions, and FOC is the flare-out condition.
\begin{table}[!htb]
\centering
\resizebox{.9\columnwidth}{!}{
\begin{tabular}{|c|c|c|c|c|}
\hline
NEC& WEC& SEC& DEC& FOC\\
\hline
 $P+\rho\geq0$& $\rho\geq0 \text{ and NEC}$& $\rho+3P\geq 0 \text{ and NEC}$& $\rho\geq |P|$& $\dot{b}(r)<1$\\
\hline
\end{tabular}}
\caption{Energy and flare-out condition.}
\label{tab-conditions}
\end{table}

In order to bound the $w$ parameter, lets us assume that $\rho\geq0$. Once that $P= w \rho$ the DEC holds for $\rho\geq |w\rho|$ so $ -1\leq w \leq 1$. On the other hand, the SEC is verified for $\left(1+3w\right)\rho\geq 0$ so  $w\leq -\frac{1}{3}$. The NEC is expressed as $(1+w)\rho\geq0$, so $w\geq-1$. Hence in order to obey NEC, SEC and DEC, the $w$ parameter is must be such that
\begin{equation}
-\frac{1}{3}< w \leq 1. \label{cond-w}
\end{equation}

\begin{figure}[!htb]
\centering
\includegraphics[scale=0.45]{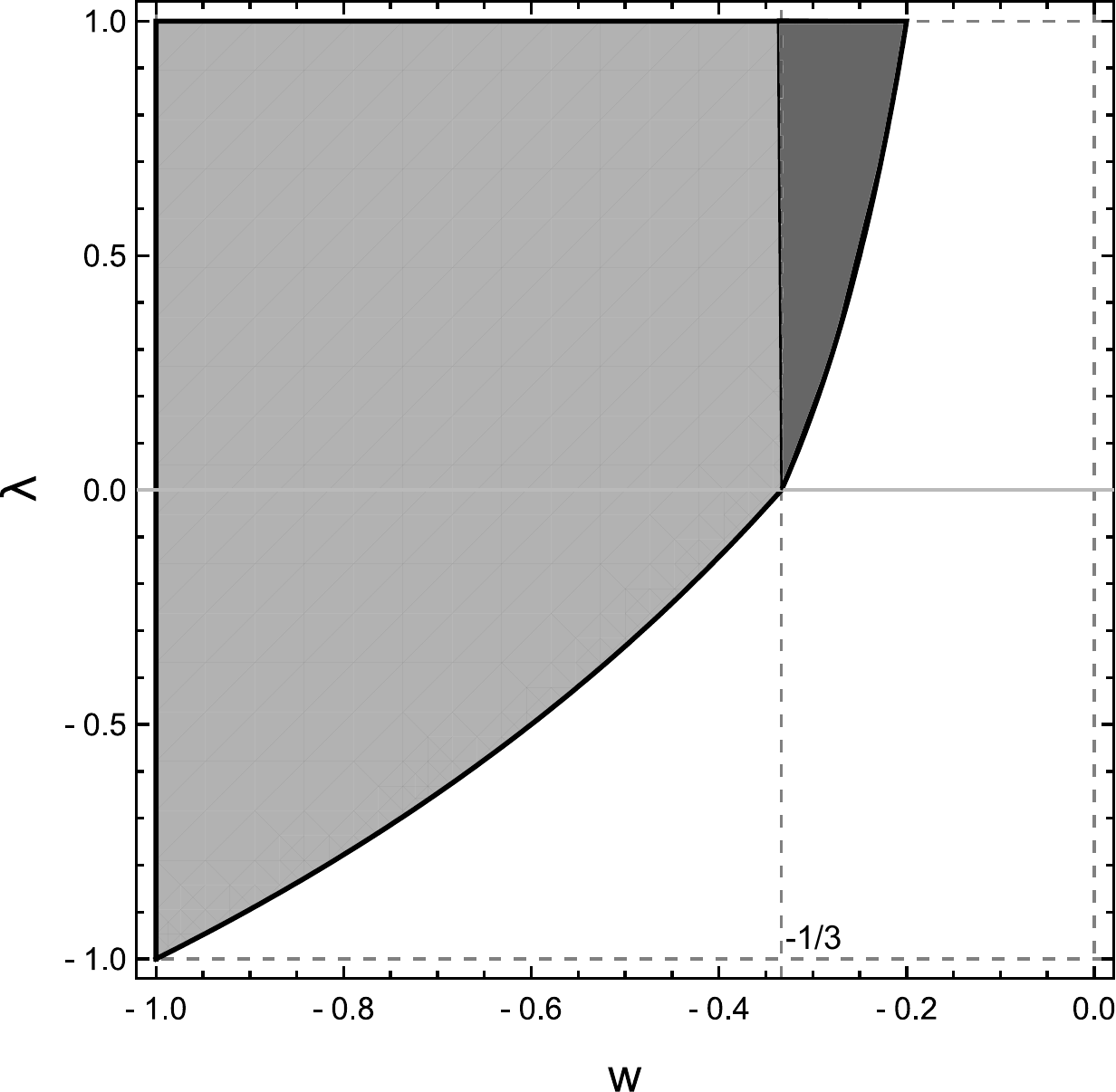}
\caption{Representation of the region where the energy conditions are satisfied. In the dark region, all energy conditions hold. In the gray region,  only the WEC is secured.}
\label{fig-lw}
\end{figure}

From now on, we set $\kappa=r_0=1$. The energy \eqref{rho-b} is positive when $\lambda> \frac{3w+1}{w-1}$ and it vanishes when $r\to\infty$ for $\gamma>3$. The region where the conditions are valid is showed in Fig. \ref{fig-lw}, being all energy condition satisfied in the dark region.

On the other hand, the flare-out condition (FOC) is necessary to maintain the structure of the wormhole traversable \cite{Ovgun, flare, morris2, zang}. The FOC can be written as 
\begin{equation}
b(r) - r \leq 0 \quad \text{and} \quad r\, \dot{b}(r) - b(r) < 0 \quad \Rightarrow \quad \dot{b}(r) < 1. \label{cond-f}
\end{equation}
Considering the equation \eqref{dbr}, the FOC is satisfied for  $\gamma>0$ if $\lambda>0$ and $r>r_0$. For the vanishing of energy \eqref{rho-b} when $r\to \infty$, it is necessary that $\gamma>-3$. So, the FOC is always obeyed when the energy \eqref{rho-b} goes to zero at infinity.

\subsection{Tortoise coordinates and the energy and flare-out conditions}\label{sec-tort}

In section \ref{sec-rw}, we will obtain the Regge-Wheeler equation for the \"{O}vg\"{u}n bumblebee wormhole.  For this goal, a  transformation to the radial variable $r$ into a new variable $x$ (tortoise coordinate) is required. This transformation, for $\Lambda=0$, is given by following the integral\cite{berg,kim}
\begin{equation}
 x = \int dr \frac{1}{\sqrt{1- \frac{b(r)}{r}}}. \label{eq-x}
 \end{equation}

To obtain an analytical function into $x$ coordinate, we choose $w=-1$ and $r_0=1$, which transform the shape-function of Eq. \eqref{br} into
\begin{equation}
b(r)=\alpha r + \frac{\beta}{r}, \label{br-1}
\end{equation}
where $\alpha=\frac{\lambda}{\lambda+1}$ and $\beta=\frac{1}{\lambda+1}=\left(1-\alpha\right)$.

Hence, by solving Eq. \eqref{eq-x} with the of Eq. \eqref{br-1}, the tortoise coordinate can be expressed as
 \begin{equation}
 x = \beta^{-1}\sqrt{\beta\left(r^2-1\right)} \quad  \Rightarrow \quad r=\sqrt{\beta x^2+1}. \label{x-r}
 \end{equation}
 
\begin{figure}[!htb]
\centering
\includegraphics[scale=0.45]{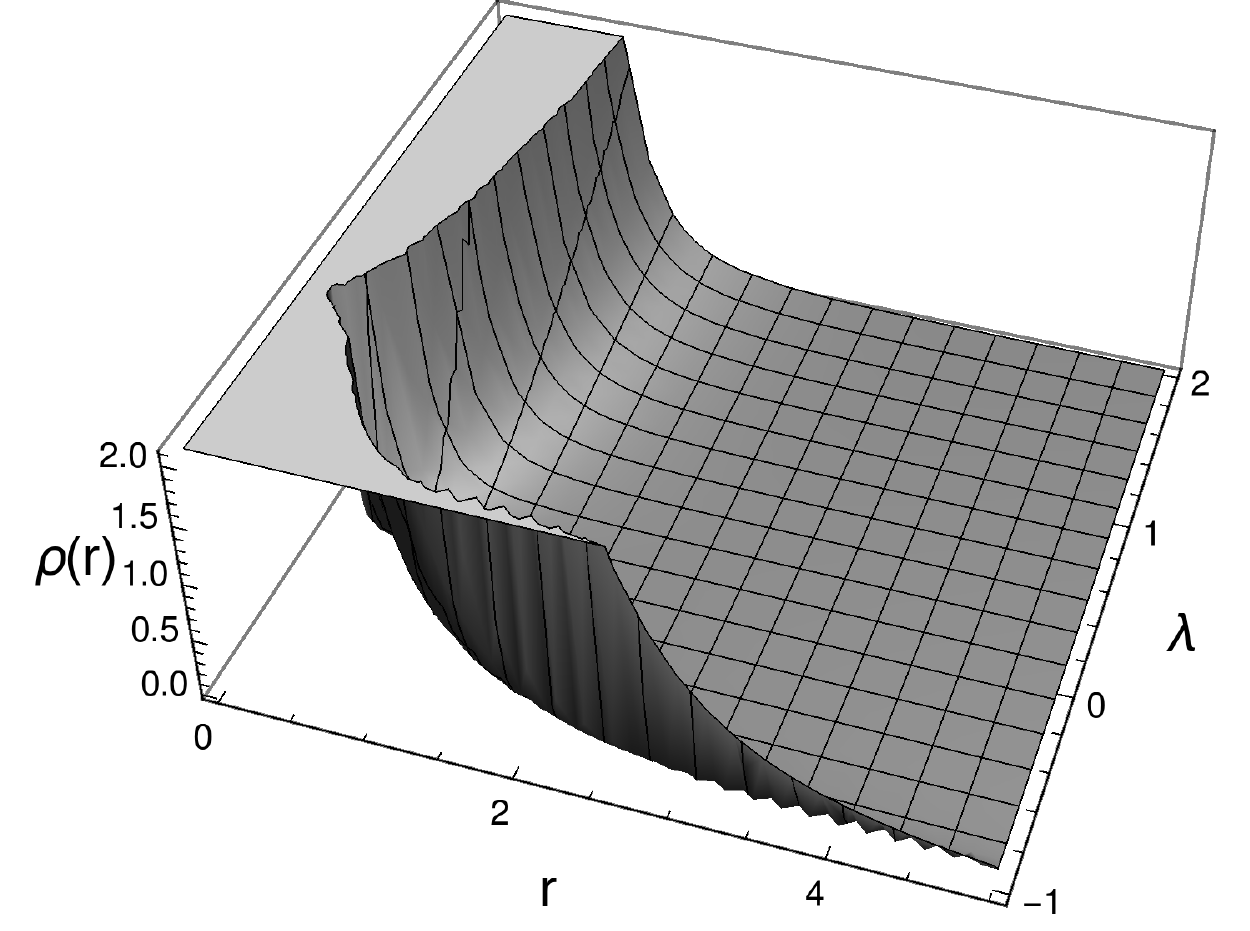}
\includegraphics[scale=0.45]{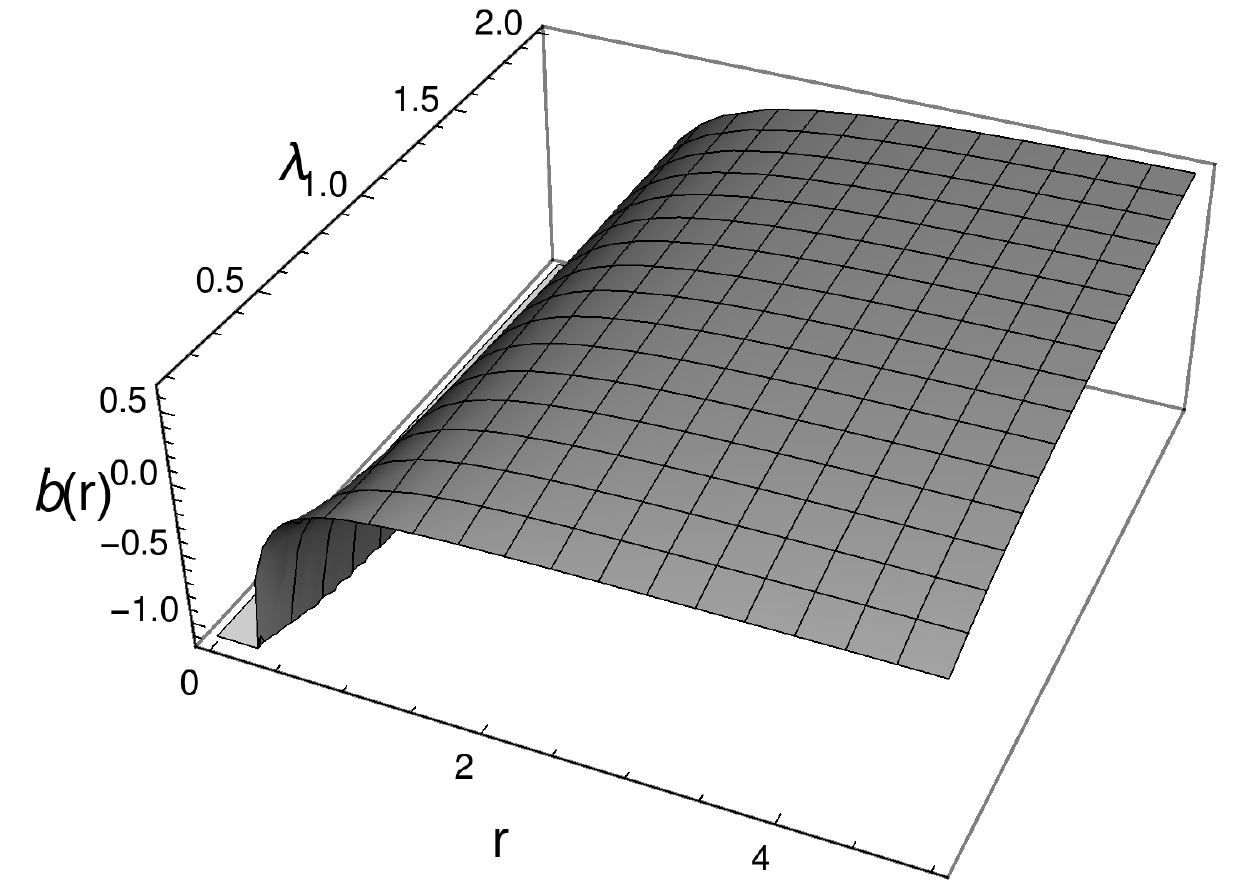}
\caption{The energy density $\rho(r)$ is always positive for $w=-1$ and $\lambda>1$ (left panel). The flare-out conditions is always verified for $w=-1$ and $\lambda>0$ (right panel).}
\label{fig-3d}
\end{figure}

For this particular choice of $w=-1$, the SEC is violated. However, the energy of Eq. \eqref{rho-b} becomes $\rho(r,\lambda)=\frac{\lambda^2}{\lambda+1}r^{-4}$ which is positive when $\lambda>-1$, as can be seen in Fig. \ref{fig-3d}. So the DEC, NEC and WEC still hold. The FOC obtained by $\dot{b}(r)$ of Eq. \eqref{br-1} reads $\frac{\lambda}{(\lambda +1)}-\frac{r^{-2}}{(\lambda +1)}<1$, which it is still valid for $\lambda>-1$, as also can be seen in Fig. \ref{fig-3d}. 
 
\begin{figure}[!htb]
\centering
\includegraphics[scale=0.55]{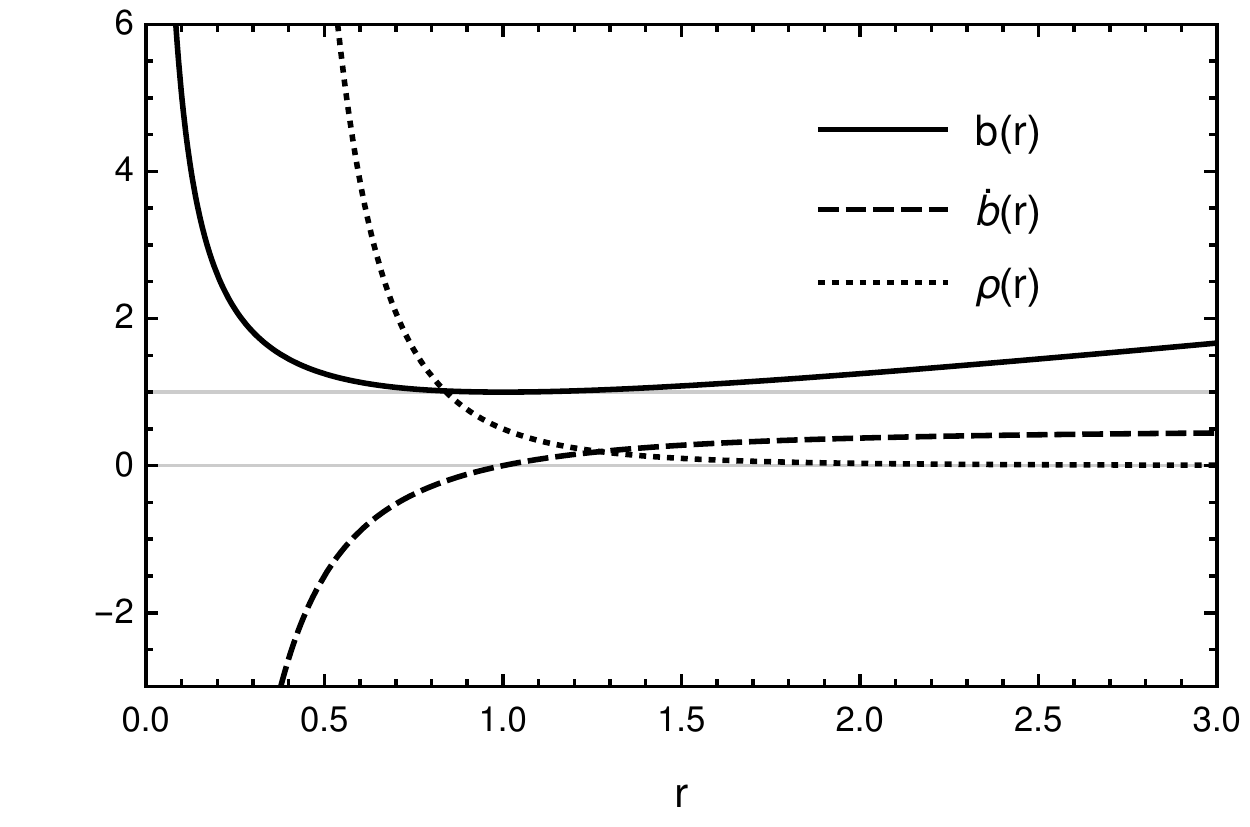}
\caption{Plot of the shape-function $b(r)$, its derivative $\dot{b}(r)$ and the energy density $\rho(r)$,  with $w=-1$ and $\lambda=1$. The gray straight lines  $y=0$ and $y=1$ help us to verify the WEC and the FOC, respectively.}
\label{fig-3p}
\end{figure}

Furthermore, in the section \ref{sec-rw} we need to set a value for the $\lambda$ to compute the quasi-normal modes. For a qualitative analysis, let us choose $\lambda=1$. The Fig. \ref{fig-3p} shows the energy density $\rho(r)$,  the shape-function $b(r)$ and its derivative $\dot{b}(r)$ with $w=-1$ and $\lambda=1$ fixed. Therefore, we note that, except the SEC, all other conditions above mentioned are satisfied.

\section{Regge-Wheeler equation and quasi-normal modes}\label{sec-rw}

In this section, we obtain the wave equation for the bumblebee wormhole. Let us consider an external perturbation, ignoring the back-reaction, and following all detailed linearization procedure for the scalar and the axial gravitational perturbations of the references \cite{berg,kim, kim2,kim3, victor}. By considering a stationary solutions in the form $\Psi(x,t)=\varphi(x)\e^{-i\omega t}$, a simplified version of the so-called Regge-Wheeler equation can be represented by \cite{regge, zerilli,  tensor2} 
\begin{equation}
\left( \frac{d^{2}}{dx^2} + \omega^2 - V(r, l)\right) \varphi(x) = 0, \label{rw-eq}
\end{equation}
where $\varphi(x)$ is the wave function (for the scalar or gravitational perturbations) with $x$ tortoise coordinate \eqref{x-r}, $\omega$ is the frequency and $V(r, l)$ is the Regge-Wheeler potential where $l$ is the azimuthal quantum number, related to the angular momentum.

For the scalar perturbations, the Regge-Wheeler potential $V(r, l)$ reads \cite{kim, kim3}
\begin{equation}
V_s(r, l) = e^{2\Lambda}\left[\frac{l\left( l + 1\right) }{r^2} - \frac{\dot{b}r - b}{2r^3} + \frac{1}{r}\left(1 - \frac{b}{r} \right)\dot{\Lambda}\right].
\label{pot-scalar}
\end{equation}
Since that $\Lambda=0$, by substituting the $b(r)$ \eqref{br-1} into potential \eqref{pot-scalar},  the $V_s(r, l) = \frac{l(l + 1)}{r^2} + \frac{\beta}{r^4}$ into $x$ coordinate \eqref{x-r} becomes
\begin{equation}
\quad V_s(x,l) = \frac{l(l + 1)}{\left(\frac{x^2}{\lambda +1}+1\right)} + \left(\frac{1}{ \lambda +1}\right)\left(\frac{x^2}{\lambda +1}+1\right)^{-2}. \label{v-scalar-1}
\end{equation}

Similarly, for the axial gravitational perturbations, the Regge-Wheeler potential $V(r, l)$ reads \cite{chan, bron, kim2, tensor2, kim3}
\begin{equation}
V_g(r, l) = e^{2\Lambda}\left[\frac{l\left( l + 1\right) }{r^2} + \frac{\dot{b}r - 5b}{2r^3} + \frac{1}{r}\left(1 - \frac{b}{r} \right)\dot{\Lambda} \right].
\end{equation}
With $b(r)$ given by Eq. \eqref{br-1}, the $V_g(r, l) = \frac{l(l + 1)- 2\alpha}{r^2} - \frac{3\beta}{r^4}
$ into $x$ coordinate \eqref{x-r} becomes
\begin{equation}
V_g(x,l) = \left(l(l + 1)- 2\frac{\lambda}{\lambda+1}\right)\left(\frac{x^2}{\lambda+1}+1\right)^{-1} - \frac{3}{\lambda+1}\left(\frac{x^2}{\lambda+1}+1\right)^{-2}\, .
\label{v-gravi-1}
\end{equation}

\begin{figure}[!htb]
\centering
\mbox{\subfigure{\includegraphics[scale=0.40]{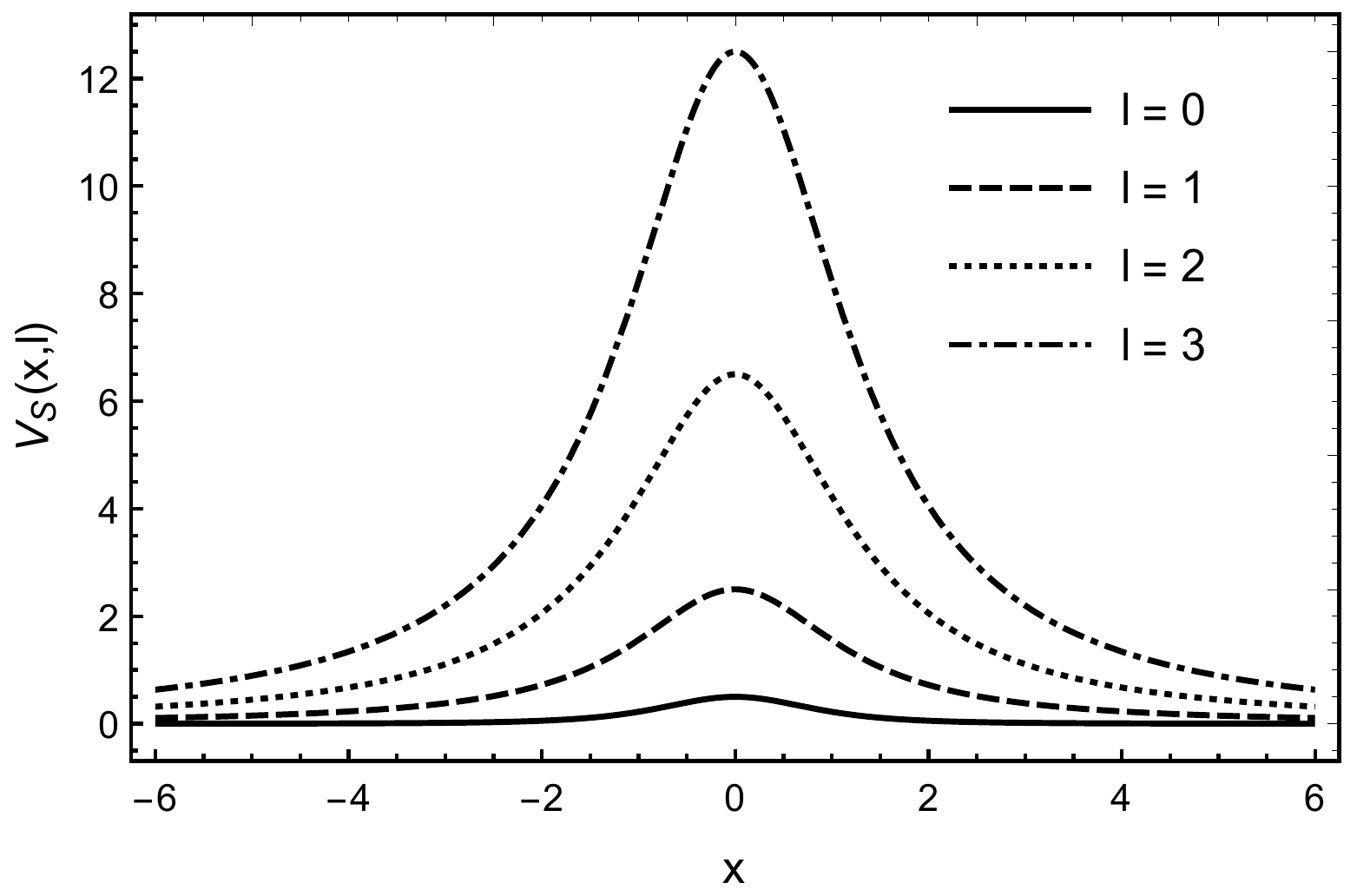}}\, 
\subfigure{\includegraphics[scale=0.40]{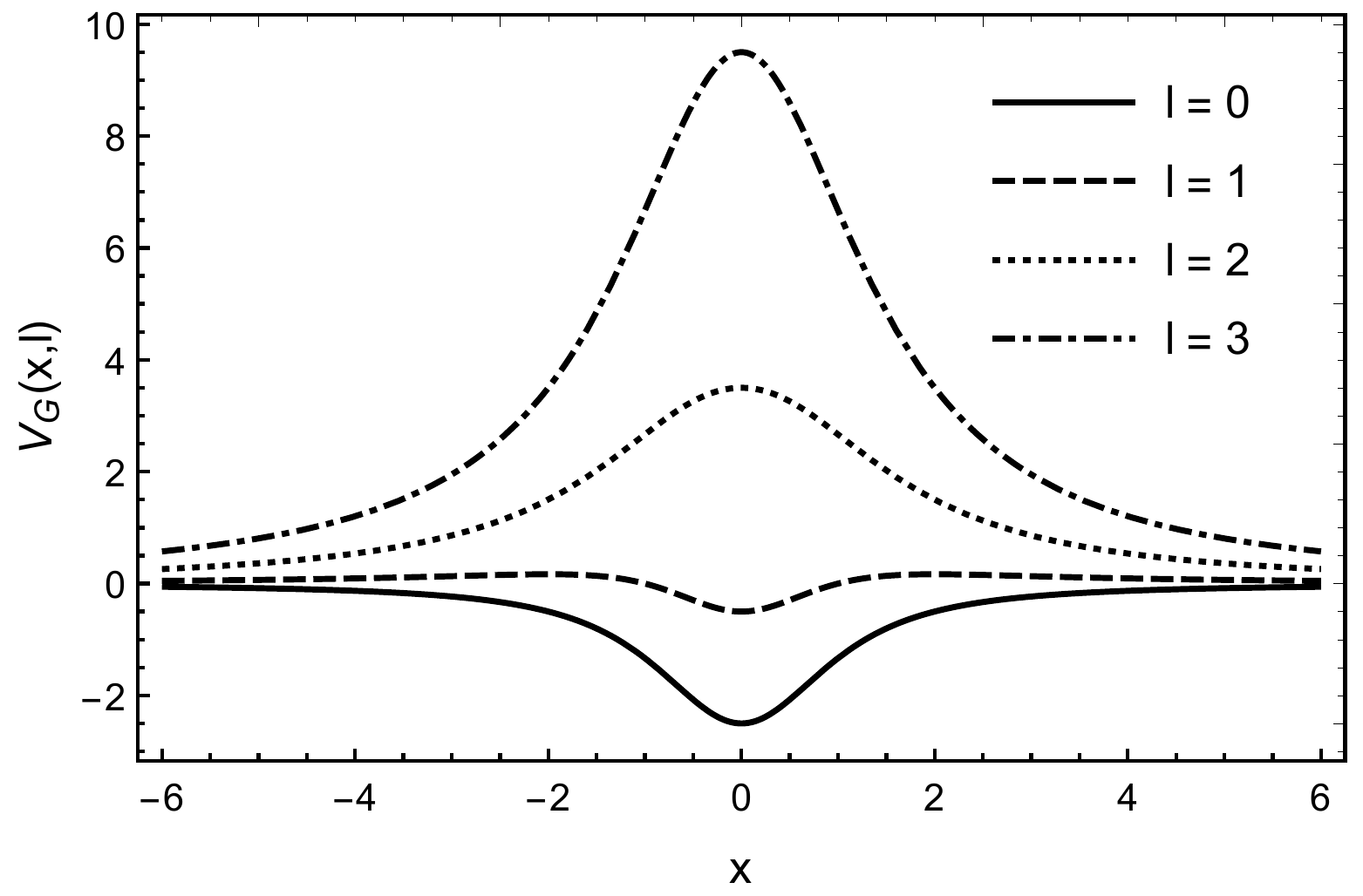}}}
\caption{The scalar (left) and gravitational (right) potentials of Regge-Wheeler equation for some $l$ parameters and $\lambda=1$.}
\label{pots}
\end{figure}

The Fig. \ref{pots} shows the plots of the scalar potential of Eq. \eqref{v-scalar-1} and gravitational potential of Eq. \eqref{v-gravi-1}. Note that both potentials are symmetric bell-shaped potentials centered at the origin. The increasing of the angular momentum increases the peaks of potentials. For the scalar perturbation all potentials are repulsive. However, for the tensorial perturbation the first two values of $l$ lead to attractive potentials. The height of peaks changes the behavior of quasi-normal modes, as will be discussed in the next subsection.

\subsection{Quasi-normal modes and the time-domain}\label{sec-qnm}

To compute the quasi-normal modes (QNMs) of Regge-Wheeler equation \eqref{rw-eq} we apply the semianalytic method of the third-order WKB approximation presented in Ref. \cite{iyer}. This method requires a positive bell-shaped potential \cite{iyer}. Hence we impose $l\geq 2$ for the tensorial perturbations.

Briefly, the QNMs can be found from the formula \cite{iyer, kono, victor}
\begin{equation}
\label{eq-qnm-formula}
\frac{\text{i}(\omega_n^2 - V_0)}{\sqrt{-2\ddot{V}_0}} - \sum_{i=2}^{6}\Lambda_{i} =
n+\frac{1}{2},
\end{equation}
where $\ddot{V}_{0}$ is the second derivative of the potential on the maximum $x_{0}$,  $\Lambda_i$ are constant coefficients, and $n$ denotes the  number of modes.

As a result, the quasi-normal modes for the scalar perturbations are written in Table \ref{tab-s}. Note that for lower $l$, larger $n$ can be unstable (denoted by \textemdash, which were excluded due the positive imaginary part). This fact can be correlated to the small peaks of the scalar potential in Fig. \ref{pots}.

\begin{table}[!htb]
%\scalefont{0.8}
\centering
\resizebox{\columnwidth}{!}{%
\begin{tabular}{| c | c | c | c |  c |}
\hline
  $\omega_n$ & $ l = 0$ & $ l = 1$ & $l = 2$ & $l = 3$ \\ \hline
 $n = 0$ & $\pm 0.428623 - 0.496200\,  i$ &$\pm  1.481593 - 0.364167  \, i$ & $\pm 2.495248 - 0.357844 \, i$ & $\pm 3.498181 - 0.355942 \, i$  \\
 $n = 1$ & $\pm 0.335244 - 1.557205\, i$ &$\pm  1.283774 - 1.155671  \, i$& $\pm 2.382690 - 1.089103 \, i$ & $\pm  3.421610 - 1.074500 \, i$  \\
 $n = 2$ & $\pm 0.184733 - 2.616606\, i$ &$\pm  0.990627 - 2.041774 \, i$ & $\pm 2.175852 - 1.859450  \, i$ & $\pm 3.274020 - 1.811550   \, i$  \\ 
 $n = 3$ &\textemdash &$\pm 0.634513 - 2.973230  \, i$ & $\pm  1.898997 - 2.674104\, i $ & $\pm 3.064840 - 2.574650  \, i$   \\ 
 $n = 4$ & \textemdash &$\pm  0.204691 - 3.93439  \, i$ & $\pm  1.56697 - 3.52542\, i $ & $\pm 2.80408 - 3.36589  \, i$   \\ 
  $n = 5$ & \textemdash & \textemdash & $\pm  1.18406 - 4.40578\, i $ & $\pm 2.49918 - 4.18339   \, i$   \\ 
 \hline
\end{tabular}}
\caption{QNMs for the scalar perturbations \eqref{v-scalar-1} with $w=-1$, $\kappa=r_0=\lambda=1$ and some $l$.}
\label{tab-s}. 
\end{table}

Similarly, the quasi-normal modes for the gravitational perturbations are written in Table \ref{tab-g}. We start with $l=3$, to guarantee positive potentials. Note that all quasi-normal modes are stable presenting no positive imaginary part. 

\begin{table}[!htb]
%\scalefont{0.8}
\centering
\resizebox{.8\columnwidth}{!}{%
\begin{tabular}{| c | c | c | c |}
\hline
 $\omega_n$ & $ l = 3$ & $ l = 4$ & $l = 5$ \\ \hline
 $n = 0$&  $\pm 3.04939 - 0.321674  \, i$  &  $\pm 4.156050 - 0.335541 \, i$ & $\pm  5.221450 - 0.341988  \, i$ \\
 $n = 1$&  $\pm  2.987810 - 0.974895 \, i$&   $\pm  4.101510 - 1.010760   \, i$ & $\pm  5.175990 - 1.028450  \, i$ \\
 $n = 2$&  $\pm 2.876140 - 1.653210  \, i$ &  $\pm  3.995510 - 1.697660   \, i$ & $\pm  5.086450 - 1.722130 \, i$ \\ 
 $n = 3$&  $\pm 2.730760 - 2.360980  \, i$ &  $\pm  3.843440 - 2.401700  \, i$ & $\pm  4.955440 - 2.427030 \, i$ \\ 
 $n = 4$&  $\pm 2.563960 - 3.092900  \, i$ &  $\pm 3.65168 - 3.12544   \, i$ & $\pm  4.78639 - 3.146020   \, i$ \\ 
 $n = 5$&  $\pm 2.380520 - 3.840940   \, i$ & $\pm 3.425930 - 3.868780   \, i$ & $\pm 4.583050 - 3.880670 \, i$ \\ 
 \hline
\end{tabular}}
\caption{QNMs for the gravitational perturbations \eqref{v-gravi-1} with $w=-1$, $\kappa=r_0=\lambda=1$ and some $l$.}
\label{tab-g}. 
\end{table}

\begin{figure}[!htb]
\centering
\mbox{
\subfigure{\includegraphics[scale=0.40]{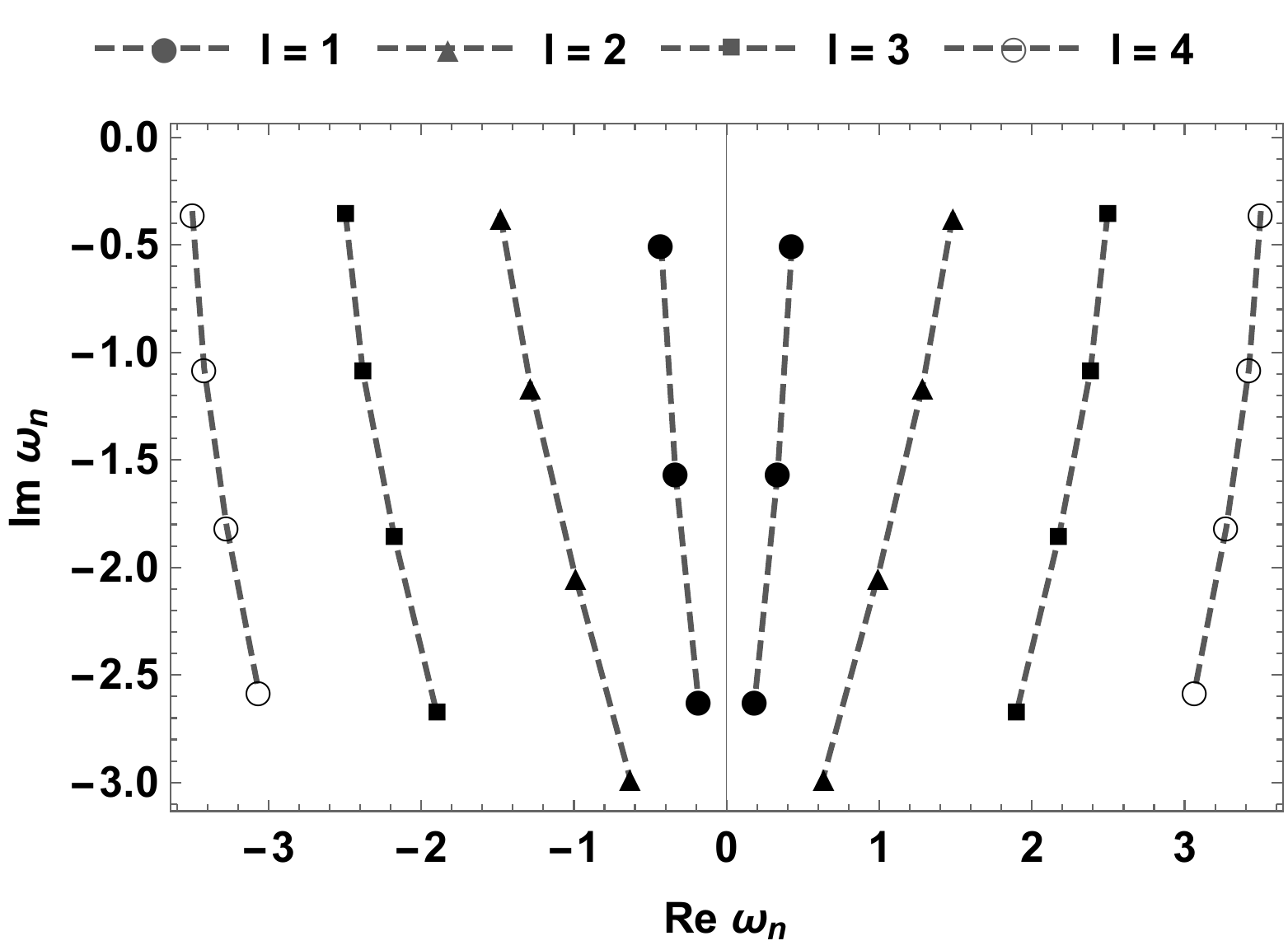}} \,
\subfigure{\includegraphics[scale=0.40]{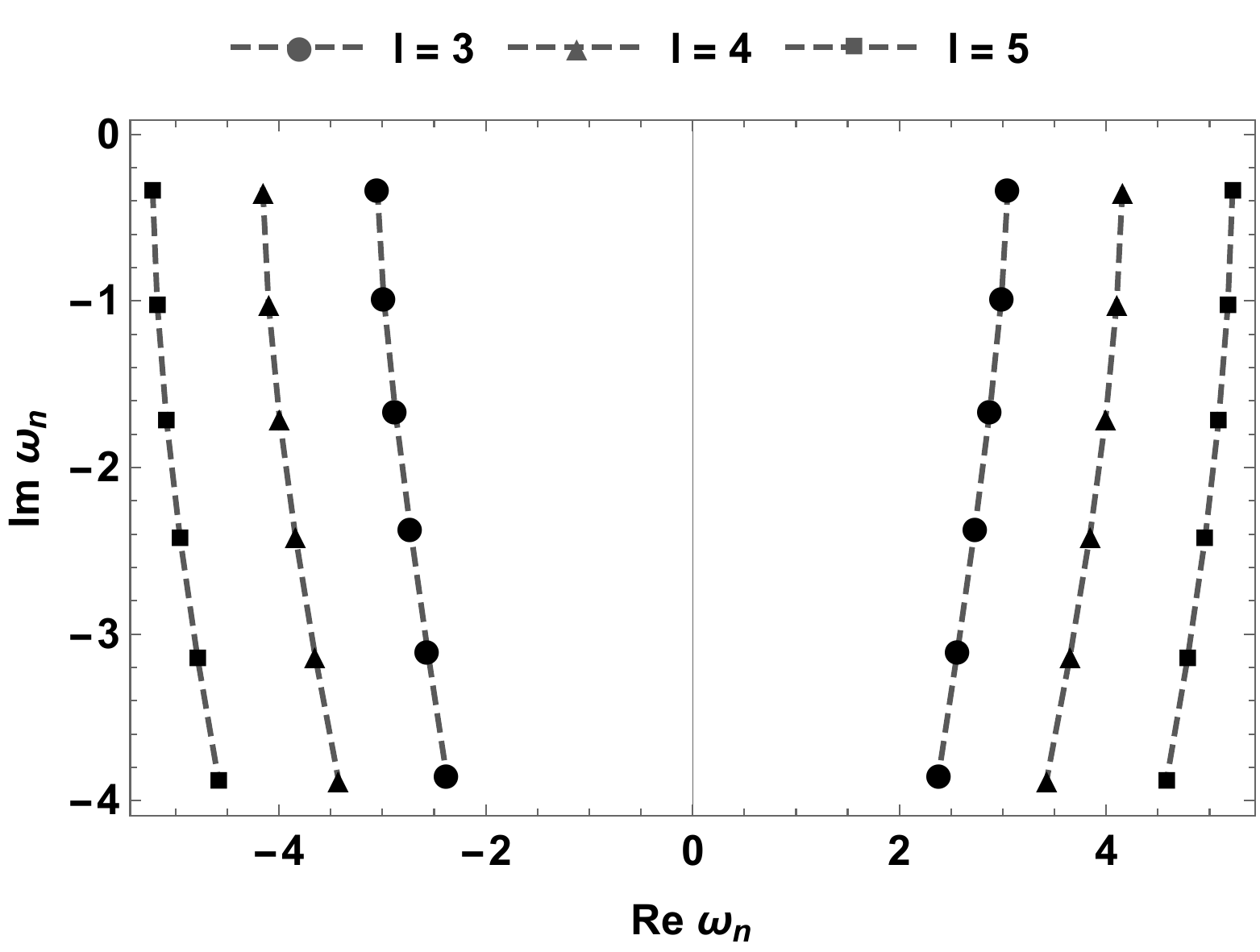}}}
\caption{Plot of QNMs for scalar (left) and gravitational (right) perturbations with $\lambda=1$.}
\label{fig-qnm}
\end{figure}

Moreover, Fig. \ref{fig-qnm} shows the points of QNMs Tables \ref{tab-s} and \ref{tab-g}. Note that all modes have smooth curves where the increase of $l$ increases the real part of each mode. The increasing of the number of modes  $n$ can leads to QNMs instabilities. 

\begin{figure}[!htb]
\centering
\mbox{
\subfigure{\includegraphics[scale=0.49]{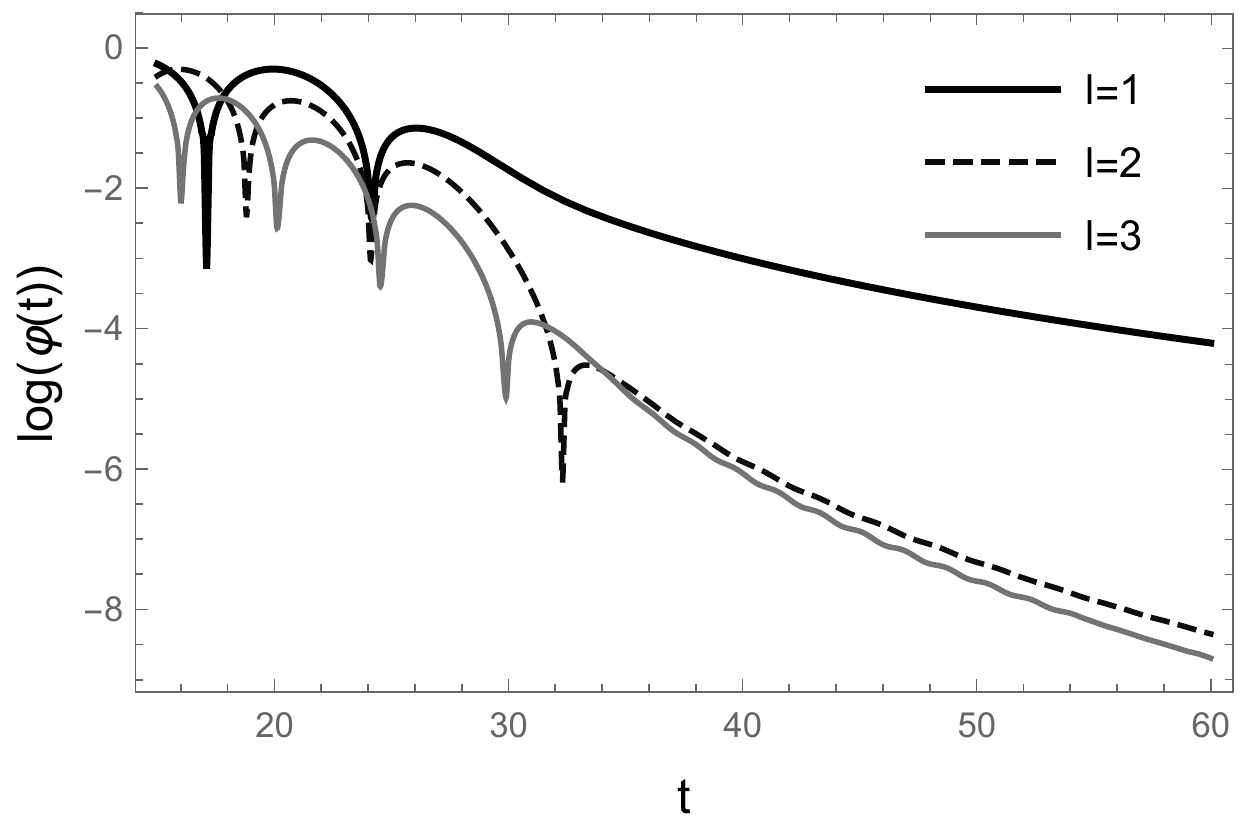}} \,
\subfigure{\includegraphics[scale=0.49]{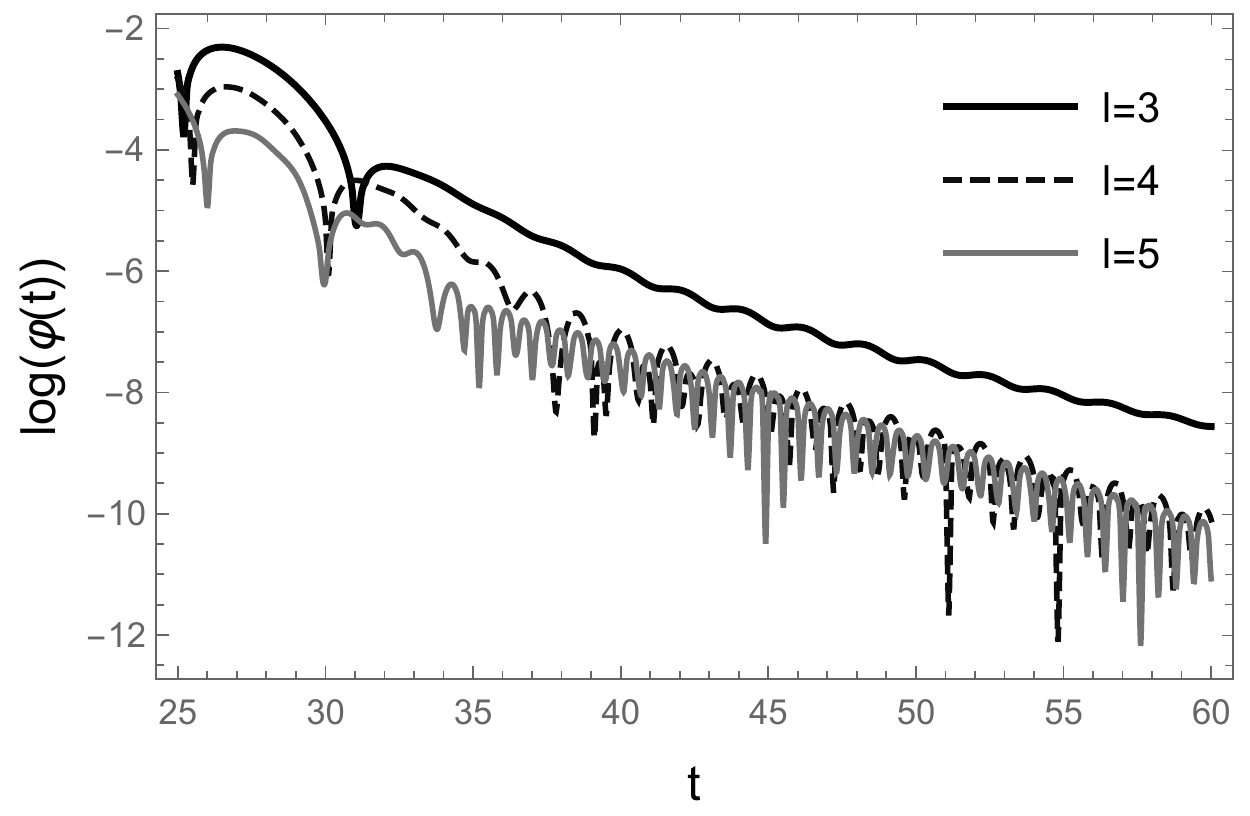}}}
\caption{Time-domain $\log_{10}(\varphi(t))$ for the scalar (left) and the gravitational (right) perturbations with $\lambda=1$.}
\label{fig-td}
\end{figure}

Besides, in Fig. \ref{fig-td}, the time-domain of bumblebee wormhole is evaluated by the Gundlach's method \cite{Gundlach}.  Note that both perturbation exhibits damping  profiles, being the decreasing of scalar modes slower than the gravitational modes.  The $l$ parameter faster the decay of these solutions.

\section{Conclusions}\label{sec-conclu}

In this work, we study the bumblebee wormhole. This scenario has a Lorentz violation parameter $\lambda$, which allows the preservation of energy conditions, leading to a wormhole generated by non-exotic matter \cite{Ovgun}. We study the possible choices of the parameters $\lambda$ and $w$ (associated to the relation $P=w\rho$) that satisfies the flare-out and the energy conditions, as shown in Fig. \ref{fig-lw}. In order to achieve an analytical and simplified tortoise coordinates transformation of Eq. \ref{x-r}, we renounce the SEC condition. However, all the other ones, namely, NEC, WEC, DEC and FOC remain valid.

Moreover, the scalar and tensorial perturbation of bumblebee wormhole were obtained. We use the general expressions for the scalar and gravitational potential. Both the potentials admit positive bell-shaped (see Fig. \ref{pots}). Hence we evaluate the quasi-normal frequencies for both perturbations. The QNMs are stable (as denoted in Tables \ref{tab-s} and \ref{tab-g}), and exhibits smooth curves (see Fig. \ref{fig-qnm}). Besides, we compute the time-domain profile for both perturbations in Fig. \ref{fig-td}, from where we note that all QNMs studied perform damping oscillation profiles.

\section*{Acknowledgments} 
%\begin{acknowledgments}
The authors thank the Funda\c{c}\~{a}o Cearense de apoio ao Desenvolvimento
Cient\'{\i}fico e Tecnol\'{o}gico (FUNCAP), the Coordena\c{c}\~{a}o de Aperfei\c{c}oamento de Pessoal de N\' ivel Superior (CAPES), and the Conselho Nacional de Desenvolvimento Cient\' ifico e Tecnol\' ogico (CNPq) for financial support.
%\end{acknowledgments}

%\section*{}


\begin{thebibliography}{99}

\bibitem{einstein-rosen} 
  A.~Einstein and N.~Rosen,
  %``The Particle Problem in the General Theory of Relativity,''
  Phys.\ Rev.\  {\bf 48}, 73 (1935).
  %doi:10.1103/PhysRev.48.73
  %%CITATION = doi:10.1103/PhysRev.48.73;%%
  %446 citations counted in INSPIRE as of 21 Nov 2018\textbf{•}


\bibitem{misner} 
  C.~W.~Misner and J.~A.~Wheeler,
  %``Classical physics as geometry: Gravitation, electromagnetism, unquantized charge, and mass as properties of curved empty space,''
  Annals Phys.\  {\bf 2}, 525 (1957).
  %doi:10.1016/0003-4916(57)90049-0
  %%CITATION = doi:10.1016/0003-4916(57)90049-0;%%
  %393 citations counted in INSPIRE as of 21 Nov 2018


\bibitem{flare} 
  D.~Hochberg and M.~Visser,
  %``Geometric structure of the generic static traversable wormhole throat,''
  Phys.\ Rev.\ D {\bf 56}, 4745 (1997).
  %doi:10.1103/PhysRevD.56.4745
  %[gr-qc/9704082].
  %%CITATION = doi:10.1103/PhysRevD.56.4745;%%
  %193 citations counted in INSPIRE as of 24 Nov 2018


\bibitem{morris} 
  M.~S.~Morris, K.~S.~Thorne and U.~Yurtsever,
  %``Wormholes, Time Machines, and the Weak Energy Condition,''
  Phys.\ Rev.\ Lett.\  {\bf 61}, 1446 (1988).
  %%doi:10.1103/PhysRevLett.61.1446 
  
\bibitem{morris2} 
  M.~S.~Morris and K.~S.~Thorne,
  %``Wormholes in space-time and their use for interstellar travel: A tool for teaching general relativity,''
  Am.\ J.\ Phys.\  {\bf 56}, 395 (1988).
  %%doi:10.1119/1.15620    
  

\bibitem{small} 
  M.~Visser, S.~Kar and N.~Dadhich,
  %``Traversable wormholes with arbitrarily small energy condition violations,''
  Phys.\ Rev.\ Lett.\  {\bf 90}, 201102 (2003).
  %%doi:10.1103/PhysRevLett.90.201102  

\bibitem{abb} 
  B.~P.~Abbott {\it et al.} [LIGO Scientific and Virgo Collaborations],
  %``Observation of Gravitational Waves from a Binary Black Hole Merger,''
  Phys.\ Rev.\ Lett.\  {\bf 116}, no. 6, 061102 (2016)  .

\bibitem{echoes} 
  P.~Bueno, P.~A.~Cano, F.~Goelen, T.~Hertog and B.~Vercnocke,
  %``Echoes of Kerr-like wormholes,''
  Phys.\ Rev.\ D {\bf 97}, no. 2, 024040 (2018).
  %doi:10.1103/PhysRevD.97.024040
  %[arXiv:1711.00391 [gr-qc]].
  %%CITATION = doi:10.1103/PhysRevD.97.024040;%%
  %26 citations counted in INSPIRE as of 22 Nov 2018

\bibitem{Energy1} 
  S.~Kar, N.~Dadhich and M.~Visser,
  %``Quantifying energy condition violations in traversable wormholes,''
  Pramana {\bf 63}, 859 (2004).
  %%doi:10.1007/BF02705207  
 

\bibitem{non-exotic} 
  R.~Shaikh,
  %``Wormholes with nonexotic matter in Born-Infeld gravity,''
  Phys.\ Rev.\ D {\bf 98}, no. 6, 064033 (2018).
  %doi:10.1103/PhysRevD.98.064033
  %[arXiv:1807.07941 [gr-qc]].
  %%CITATION = doi:10.1103/PhysRevD.98.064033;%%
  %5 citations counted in INSPIRE as of 21 Nov 2018


\bibitem{fr-wormhole} 
  N.~Godani and G.~C.~Samanta,
  ``Traversable Wormholes and Energy Conditions with Two Different Shape Functions in $f(R)$ Gravity,''
  arXiv:1809.00341 [gr-qc].
  %%CITATION = ARXIV:1809.00341;%% 

\bibitem{gb-wormhole} 
  M.~R.~Mehdizadeh, M.~K. Zangeneh and F.~S.~N.~Lobo,
  %``Einstein-Gauss-Bonnet traversable wormholes satisfying the weak energy condition,''
  Phys.\ Rev.\ D {\bf 91}, no. 8, 084004 (2015).
  %doi:10.1103/PhysRevD.91.084004
  %[arXiv:1501.04773 [gr-qc]].
  %%CITATION = doi:10.1103/PhysRevD.91.084004;%%
  %32 citations counted in INSPIRE as of 22 Nov 2018


\bibitem{qnm-wormhole} 
  S.~H.~Völkel and K.~D.~Kokkotas,
  %``Wormhole Potentials and Throats from Quasi-Normal Modes,''
  Class.\ Quant.\ Grav.\  {\bf 35}, no. 10, 105018 (2018).
  %doi:10.1088/1361-6382/aabce6
  %[arXiv:1802.08525 [gr-qc]].
  %%CITATION = doi:10.1088/1361-6382/aabce6;%%
  %12 citations counted in INSPIRE as of 21 Nov 2018

 \bibitem{Ovgun} 
  A.~Övgün, K.~Jusufi and İ.~Sakallı,
  ``Exact traversable wormhole solution in bumblebee gravity,''
  arXiv:1804.09911 [gr-qc].

%\cite{Kostelecky:1988zi}
\bibitem{kosu} 
  V.~A.~Kostelecky and S.~Samuel,
  %``Spontaneous Breaking of Lorentz Symmetry in String Theory,''
  Phys.\ Rev.\ D {\bf 39}, 683 (1989).
  %%doi:10.1103/PhysRevD.39.683
  
 
  
%\cite{Colladay:1998fq}
\bibitem{coll} 
  D.~Colladay and V.~A.~Kostelecky,
  %``Lorentz violating extension of the standard model,''
  Phys.\ Rev.\ D {\bf 58}, 116002 (1998).
  %%doi:10.1103/PhysRevD.58.116002



\bibitem{kost} 
  V.~A.~Kostelecky,
  %``Gravity, Lorentz violation, and the standard model,''
  Phys.\ Rev.\ D {\bf 69}, 105009 (2004).
  %%doi:10.1103/PhysRevD.69.105009
  

  

\bibitem{casana} 
  R.~Casana, A.~Cavalcante, F.~P.~Poulis and E.~B.~Santos,
  %``Exact Schwarzschild-like solution in a bumblebee gravity model,''
  Phys.\ Rev.\ D {\bf 97}, no. 10, 104001 (2018).
  %%doi:10.1103/PhysRevD.97.104001   

\bibitem{ellis} K. Jusufi, Int. J. Geom. Methods Mod. Phys. 14, 1750179 (2017).

\bibitem{regge} 
  T.~Regge and J.~A.~Wheeler,
  %``Stability of a Schwarzschild singularity,''
  Phys.\ Rev.\  {\bf 108}, 1063 (1957).
  
\bibitem{iyer} 
  S.~Iyer and C.~M.~Will,
  %``Black Hole Normal Modes: A {WKB} Approach. 1. Foundations and Application of a Higher Order {WKB} Analysis of Potential Barrier Scattering,''
  Phys.\ Rev.\ D {\bf 35}, 3621 (1987).
  %%doi:10.1103/PhysRevD.35.3621  
  
%\bibitem{hawking} Hawking, S., \& Ellis, G. (1973). \textit{The Large Scale Structure of Space-Time} (Cambridge Monographs on Mathematical Physics). Cambridge: Cambridge University Press. %%doi:10.1017/CBO9780511524646  

\bibitem{zang} 
  M.~Kord Zangeneh, F.~S.~N.~Lobo and M.~H.~Dehghani,
  %``Traversable wormholes satisfying the weak energy condition in third-order Lovelock gravity,''
  Phys.\ Rev.\ D {\bf 92}, no. 12, 124049 (2015).
  %%doi:10.1103/PhysRevD.92.124049
  
\bibitem{kim} 
  S.~W.~Kim,
  %``Rotating wormhole and scalar perturbation,''
  Nuovo Cim.\ B {\bf 120}, 1235 (2005).
  
\bibitem{berg} 
  S.~E.~Perez Bergliaffa and K.~E.~Hibberd,
  %``Electromagnetic waves in a wormhole geometry,''
  Phys.\ Rev.\ D {\bf 62}, 044045 (2000).

\bibitem{kim2} 
  S.~W.~Kim,
  ``Gravitational perturbation of traversable wormhole,''
  gr-qc/0401007.

%\cite{Kim:2008zzj}
\bibitem{kim3} 
  S.~W.~Kim,
  %``Wormhole perturbation and its quasi normal modes,''
  Prog.\ Theor.\ Phys.\ Suppl.\  {\bf 172}, 21 (2008).
  %doi:10.1143/PTPS.172.21
  %%CITATION = doi:10.1143/PTPS.172.21;%%
  %2 citations counted in INSPIRE as of 25 Nov 2018
  
\bibitem{victor} 
  V.~Santos, R.~V.~Maluf and C.~A.~S.~Almeida,
  %``Quasinormal frequencies of self-dual black holes,''
  Phys.\ Rev.\ D {\bf 93}, no. 8, 084047 (2016)
  %doi:10.1103/PhysRevD.93.084047
  %[arXiv:1509.04306 [gr-qc]].
  %%CITATION = doi:10.1103/PhysRevD.93.084047;%%
  %11 citations counted in INSPIRE as of 25 Nov 2018



\bibitem{zerilli} 
  F.~J.~Zerilli,
  %``Effective potential for even parity Regge-Wheeler gravitational perturbation equations,''
  Phys.\ Rev.\ Lett.\  {\bf 24}, 737 (1970).
  %%doi:10.1103/PhysRevLett.24.737   

\bibitem{tensor2} 
  P.~Boonserm, T.~Ngampitipan and M.~Visser,
  %``Regge-Wheeler equation, linear stability, and greybody factors for dirty black holes,''
  Phys.\ Rev.\ D {\bf 88}, 041502 (2013)
  %doi:10.1103/PhysRevD.88.041502
  %[arXiv:1305.1416 [gr-qc]].
  %%CITATION = doi:10.1103/PhysRevD.88.041502;%%
  %13 citations counted in INSPIRE as of 25 Nov 2018  

  
\bibitem{chan} 
 S.~Chandrasekhar,
  ``The mathematical theory of black holes,''
  Oxford, UK: Clarendon (1985) 646 P.
 
\bibitem{bron} 
  K.~A.~Bronnikov, R.~A.~Konoplya and A.~Zhidenko,
  %``Instabilities of wormholes and regular black holes supported by a phantom scalar field,''
  Phys.\ Rev.\ D {\bf 86}, 024028 (2012) 
  

  
  
  
\bibitem{kono} 
  R.~A.~Konoplya,
  %``Quasinormal behavior of the d-dimensional Schwarzschild black hole and higher order WKB approach,''
  Phys.\ Rev.\ D {\bf 68}, 024018 (2003) 
  
%\bibitem{gur} 
%  U.~Gürsoy, A.~Jansen and W.~van der Schee,
%  %``New dynamical instability in asymptotically anti–de Sitter spacetime,''
%  Phys.\ Rev.\ D {\bf 94}, no. 6, 061901 (2016)  
  
   
 \bibitem{Gundlach} 
  C.~Gundlach, R.~H.~Price and J.~Pullin,
  %``Late time behavior of stellar collapse and explosions: 1. Linearized perturbations,''
  Phys.\ Rev.\ D {\bf 49}, 883 (1994)
  %doi:10.1103/PhysRevD.49.883
  %[gr-qc/9307009].
  %%CITATION = doi:10.1103/PhysRevD.49.883;%%
  %288 citations counted in INSPIRE as of 25 Nov 2018 
    
  
    
  
  

  
  
  
\end{thebibliography}
\end{document}